\documentclass{article}

\usepackage{arxiv}

\usepackage[utf8]{inputenc} %
\usepackage[T1]{fontenc}    %
\usepackage{hyperref}       %
\usepackage{url}            %
\usepackage{booktabs}       %
\usepackage{amsfonts}       %
\usepackage{nicefrac}       %
\usepackage{microtype}      %
\usepackage{lipsum}
\usepackage{graphicx}
\usepackage{textcomp}
\usepackage{xspace}
\usepackage{gensymb}
\usepackage{float}
\usepackage{graphicx}
\usepackage{amsmath}
\usepackage{authblk}
\DeclareUnicodeCharacter{221E}{\ensuremath{\infty}}
\DeclareUnicodeCharacter{1D70F}{\ensuremath{\tau}}
\usepackage[style=phys,backend=bibtex]{biblatex}
\title{Models Currently Implemented in MIIND}
\author[1]{Hugh Osborne}
\author[2]{Yi Ming Lai}
\author[1]{Marc de Kamps}
\affil[1]{School of Computing, University of Leeds}
\affil[2]{School of Mathematics, University of Nottingham}

\begin{document}

\maketitle

\begin{abstract}
This is a living document that will be updated when appropriate. MIIND \cite{dekamps2008,de2019computational} is a population-level neural simulator. It is based on population density techniques, just like DIPDE \cite{dipde}. Contrary to DIPDE, MIIND is agnostic to the underlying neuron model used in its populations so any 1, 2 or 3 dimensional model can be set up with minimal effort. The resulting populations can then be grouped into large networks, e.g. the Potjans-Diesmann model \cite{potjans2014}. The MIIND website \url{http://miind.sf.net} contains training materials, and helps to set up MIIND, either by using virtual machines, a DOCKER image, or directly from source code.      
\end{abstract}

The following is a list of models which have previously been implemented. Each section gives a quick description of the model, the mathematical definition, file references for running the model in MIIND, and some images and output of the population activity. All referenced files are in the \texttt{examples/model\_archive/} directory in the MIIND repository on GitHub. Files for the 3D models are available in the archive but are only supported in an unreleased version of MIIND.
\section{Leaky, Quadratic, and Exponential Integrate and Fire}
Common 1 dimensional models for approximating neuron behaviour. The quadratic and exponential versions provide more realistic spike shapes. All have a threshold resent mechanism for avoiding the need for repolarising variables. MIIND provides excellent speed and memory benefits in comparison to direct simulation of more than 10000 neurons \cite{de2019computational}.

\begin{align}
    \tau V' = - g_{l} ( V - E_{v} ) + I \nonumber
\end{align}
\begin{align}
    \tau V' = g_{l} ( V - E_{v} ) ( V - V_{thres} ) + I \nonumber
\end{align}
\begin{align}
    \tau V' = - g_{l} ( V - E_{v} ) + \Delta exp(\frac{V - V_{thres}}{\Delta}) + I \nonumber
\end{align}

\begin{figure}[H]
  \centering
  \includegraphics[width=0.9\columnwidth]{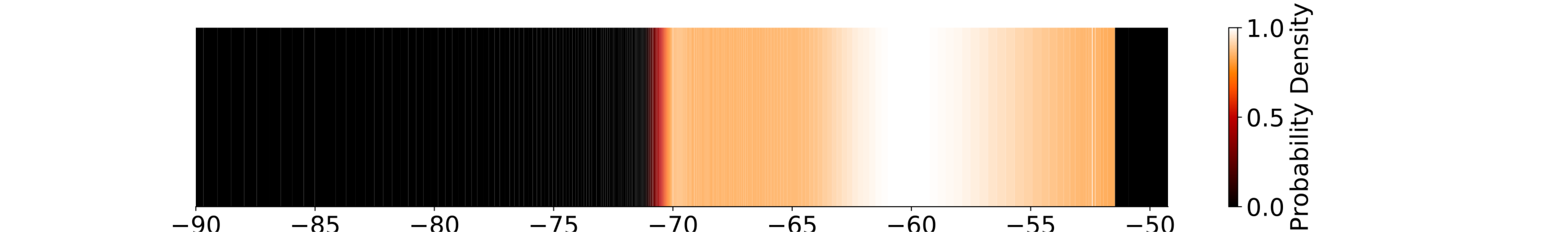}
  \caption{Density plot for a population of EIF neurons simulated in MIIND.}
  \label{fig:fig5}
\end{figure}

Files:\\
$LIF/lif.xml$\\
$LIF/lif.model$\\
$LIF/lif\_0.01\_0\_0\_0\_.mat$\\
$QIF/qif.xml$\\
$QIF/qif.model$\\
$QIF/qif\_0.01\_0\_0\_0\_.mat$\\
$EIF/eif.xml$\\
$EIF/eif.model$\\
$EIF/eif\_0.1\_0\_0\_0\_.mat$

\section{Potjans-Diesmann (cortical column)}
The original Potjans-Diesmann model \cite{potjans2014} of an 8 population model, representing a cortical column was adapted by Cain \emph{et al} \cite{cain2016}. We follow their implementation, in passing validating some of the experiments that were published on the DIPDE website \cite{dipde}.

\begin{figure}[H]
  \centering
  \includegraphics[width=0.9\columnwidth]{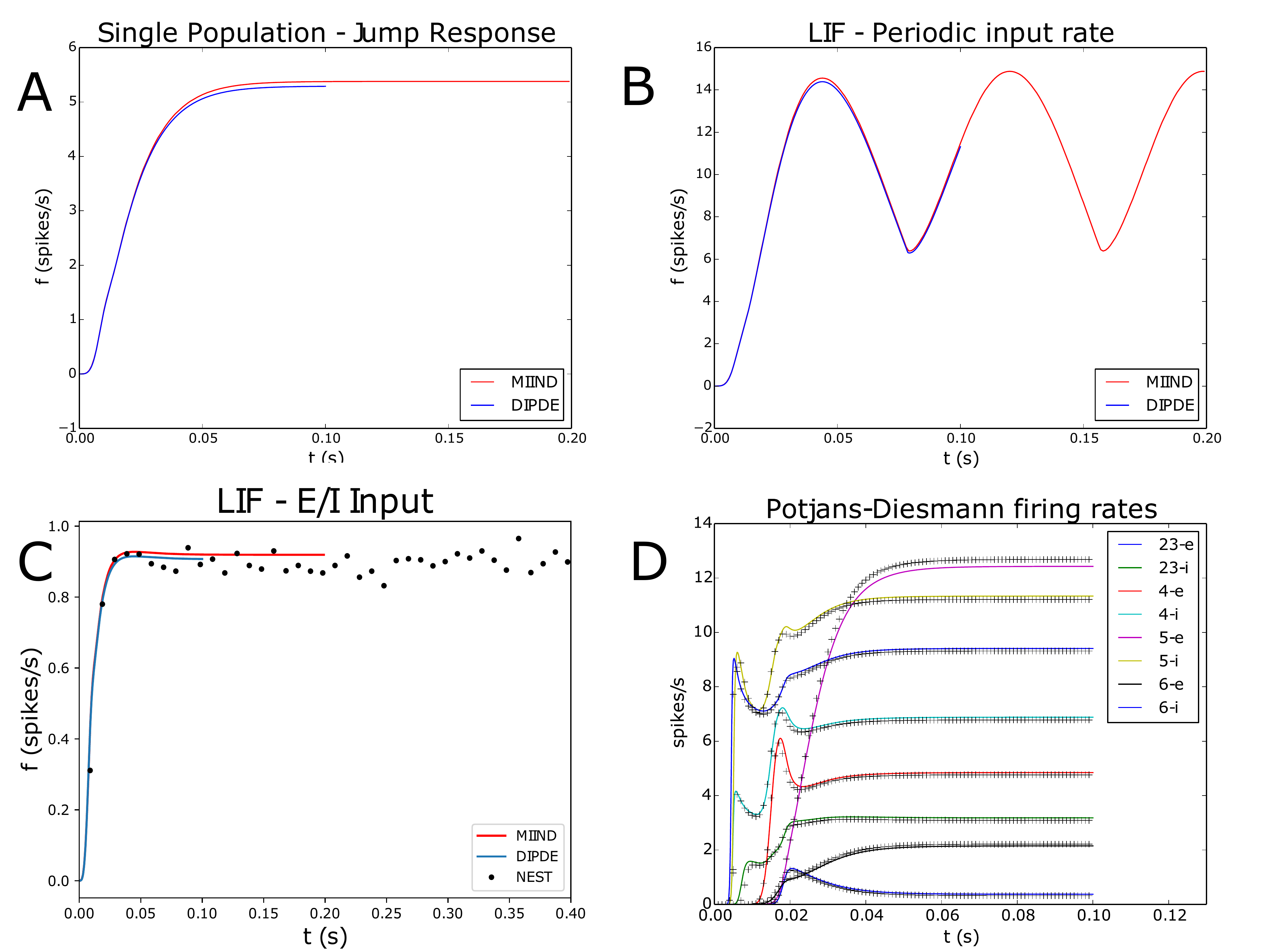}
  \caption{A: Comparison between a population receiving a single Poisson distributed spike train. B: Population receiving a combination of excitatory and inhibitory spike trains, C: excitatory input spike trains with a time dependent firing rate. D: Comparison of the Potjans-Diesmann response after being switched on at $t = 0$, both for the MIIND and the DIPDE implementation.   }
  \label{fig:figpot}
\end{figure}

In general, MIIND estimates the  steady state firing rate slightly higher (Fig. \ref{fig:figpot}), but the differences are not really significant when compared to a NEST simulation (B) for 100000 neurons. The agreement between MIIND (markers) and DIPDE (solid lines) for the full Potjans-Diesmann model is very good, both in the representation of the input dynamics, and the steady state firing rates.

Files:\\
$DIPDE/lifdipde.model$\\
$DIPDE/lifdipde_-0.005.mat$\\
$DIPDE/lifdipde_0.005.mat$ \\
$DIPDE/ei.xml$\\
$DIPDE/recurrent.xml$\\
$DIPDE/single.xml$\\
$DIPDE/variable.xml$

For the Potjans-Diesmann model:
Files:
$POTJANS/potjanslif.model$ \\
$POTJANS/potjanslif_-0.0007\_0\_0\_0\_.mat$\\	$POTJANS/potjanslif_0.000175\_0\_0\_0\_.mat$ \\
$POTJANS/potjans.xml$

\section{Adaptive Exponential Integrate and Fire (AdExp)}
\cite{brette2005adaptive} Time dependent variables are membrane potential and an adaptive variable making quick repeated firing less likely. Useful for capturing the adaptive behaviour of pyramid cells.

\begin{align}
    \tau V' = - g_{l} ( V - E_{v} ) + \Delta exp(\frac{V - V_{thres}}{\Delta}) - W + I \nonumber
\end{align}
\begin{align}
    \tau_{w} W' = \alpha ( V - E_{v} ) - W \nonumber
\end{align}

\begin{figure}[H]
  \centering
  \includegraphics[width=0.3\columnwidth]{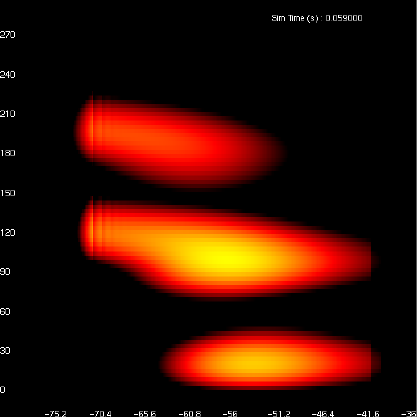}
  \caption{Density plot for a population of AdExp neurons simulated in MIIND. With a constant input rate, during the transient period, jumps in the adaptive variable can be clearly seen.}
  \label{fig:fig6}
\end{figure}

Files:\\
$Adex/adex.xml$\\
$Adex/adex.model$\\
$Adex/adex\_0\_0\_0\_0\_.tmat$

\section{Fitzhugh-Nagumo}
A two dimensional reduction \cite{nagumo1962active,fitzhugh1961impulses} of the Hodgkin-Huxley neuron.
 
\begin{align}
    V' = V - \frac{V^3}{3} - W + I \nonumber
\end{align}
\begin{align}
    W' = 0.08 ( V + 0.7 - 0.8W ) \nonumber
\end{align}

\begin{figure}[H]
  \centering
  \includegraphics[width=0.3\columnwidth]{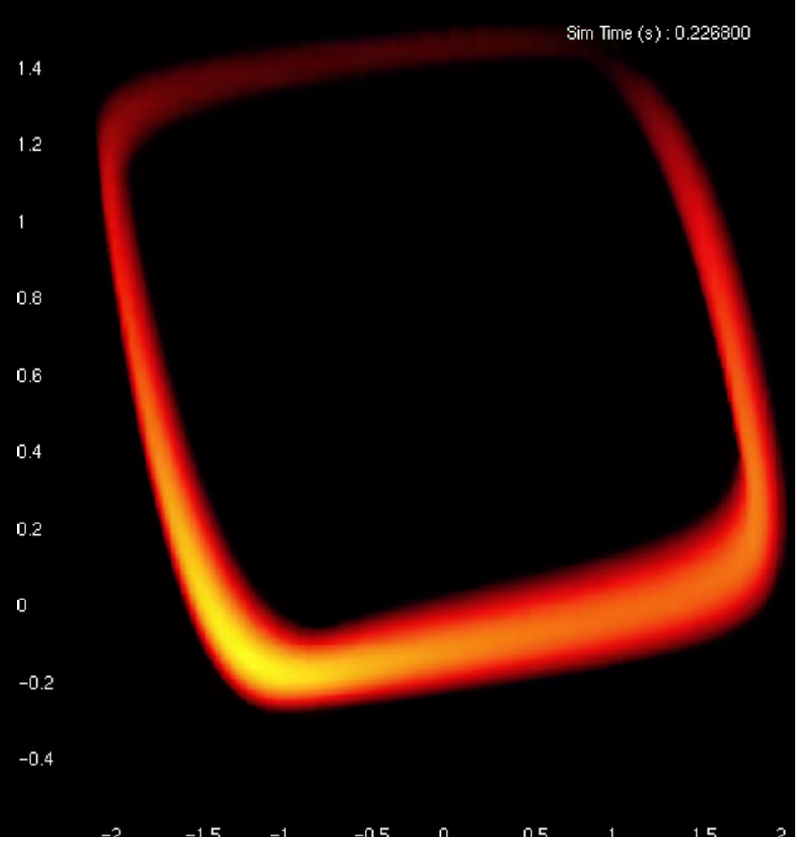}
  \includegraphics[width=0.4\columnwidth]{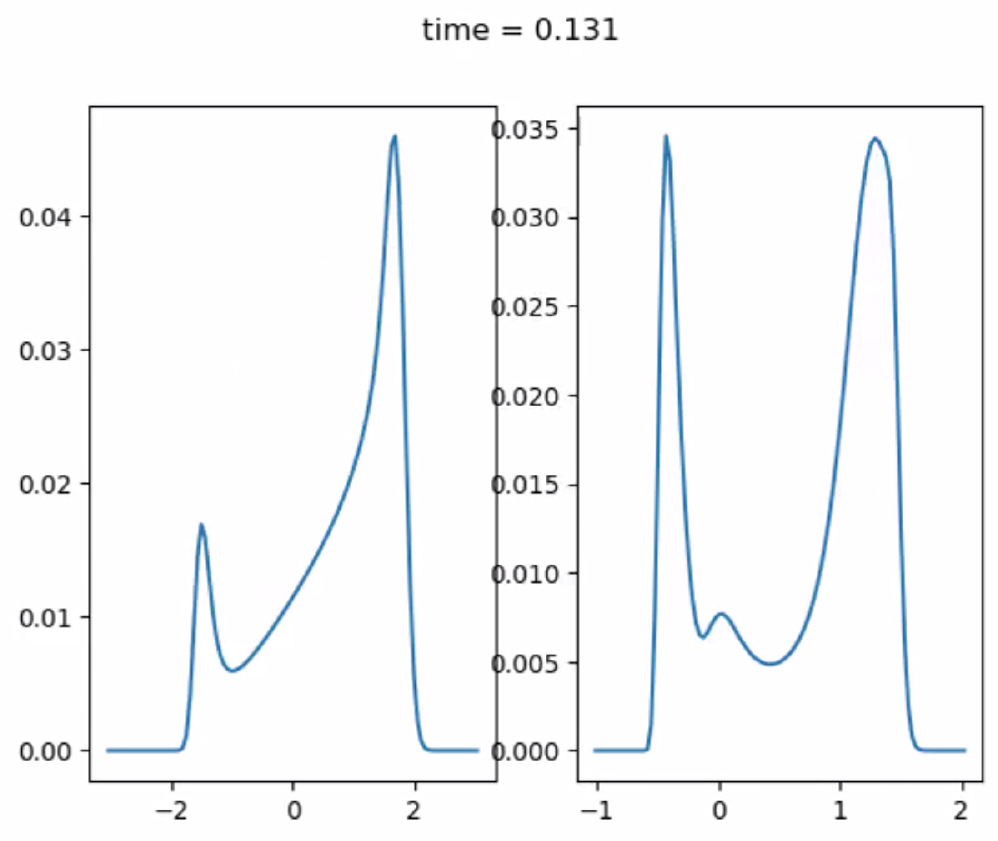}
  \caption{(left) The density plot for a population of FitzHugh-Nagumo neurons. (right) The marginal densities for the same population.}
  \label{fig:fig1}
\end{figure}

Files:\\
$FitzhughNagumo/fn.xml$\\
$FitzhughNagumo/fn.model$\\
$FitzhughNagumo/fn\_0\_0\_0\_0\_.tmat$

\section{Izhikevich Simple Model}
A simple quadratic neuron model \cite{izhikevich2003simple} parameterised ($a$,$b$,$c$, and $d$) to produce a wide range of common neuron behaviours such as bursting and fast spiking.
 
\begin{align}
    V' = 0.04V^2 + 5V + 140 - W + I \nonumber
\end{align}
\begin{align}
    W' = a ( bV - W ) \nonumber
\end{align}
\begin{align}
\text{if   } V=30mV \text{  then  } V \leftarrow c, W \leftarrow W + d
\end{align}

\begin{figure}[H]
  \centering
  \includegraphics[width=0.3\columnwidth]{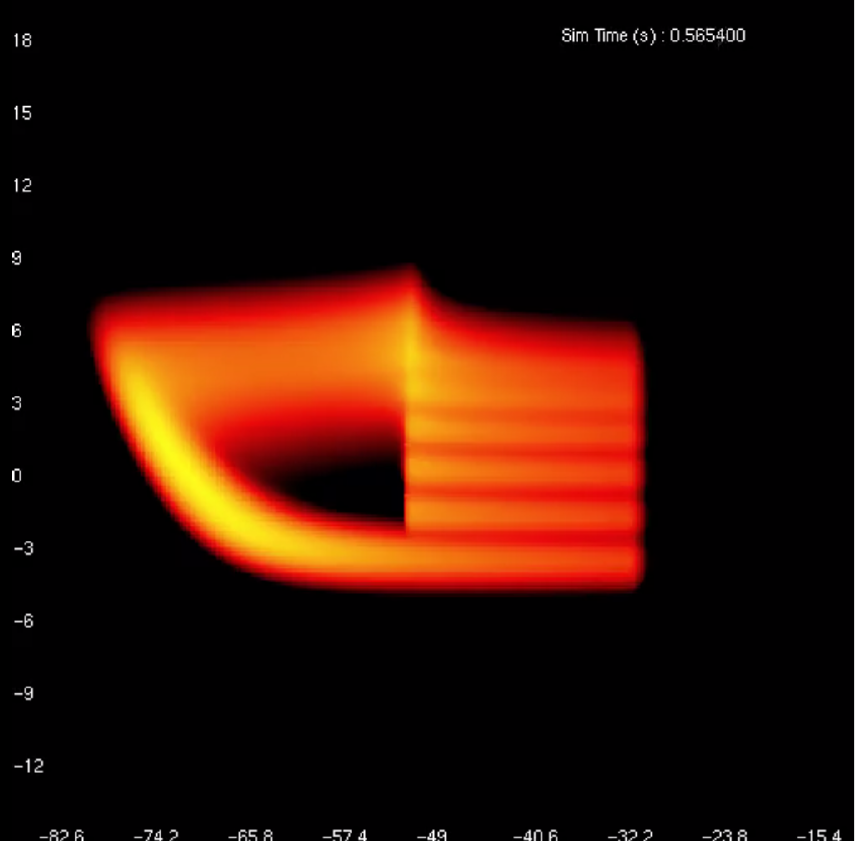}
  \includegraphics[width=0.4\columnwidth]{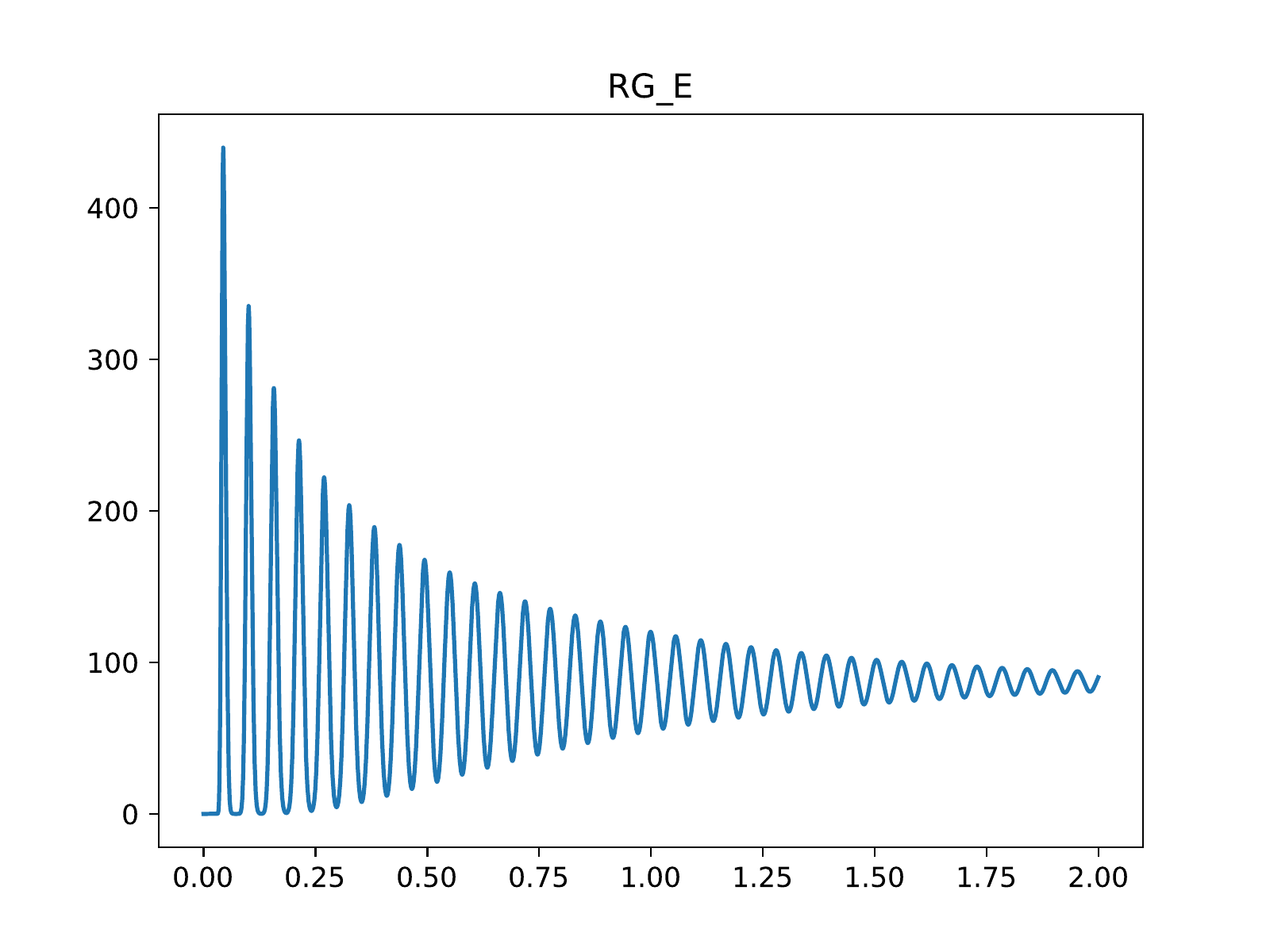}
  \caption{(left) The density plot for a population of Izhikevich neurons with their bursting parameter set. (right) The average firing rate plot for the same population.}
  \label{fig:fig2}
\end{figure}

Files:\\
$Izhikevich/izh.xml$\\
$Izhikevich/izh.model$\\
$Izhikevich/izh\_0\_0\_0\_0\_.tmat$

\section{Reduced Persistent Sodium Burster}
An HH model with additional persistent sodium channel reduced to two dimensions by removing the time dependence of the sodium and potassium gating variables. Spiking is handled with a threshold reset mechanism. Parameters as defined in \cite{rybak2015organization} to support a CPG model.
 
\begin{align}
    V' = -g_{k}m_{\infty k}^4 ( V - E_{k} ) - g_{Nap}m_{\infty Nap}h_{Nap} ( V - E_{Na} ) - g_{Na}m_{\infty Na}^3h_{\infty Na} ( V - E_{Na} ) - g_{l} ( V - E_{l} ) \nonumber
\end{align}
\begin{align}
    \tau_{Nap}h_{Nap}' = h_{\infty Nap} - h_{Nap} \nonumber
\end{align}

$g_{k}$, $g_{Nap}$, $g_{Na}$, and $g_{l}$ are constant conductances with $m_{\infty k}$, $m_{\infty Na}$,  $h_{\infty Nap}$, and $h_{\infty Na}$ representing typical sinusoidal curves.

\begin{figure}[H]
  \centering
  \includegraphics[width=0.3\columnwidth]{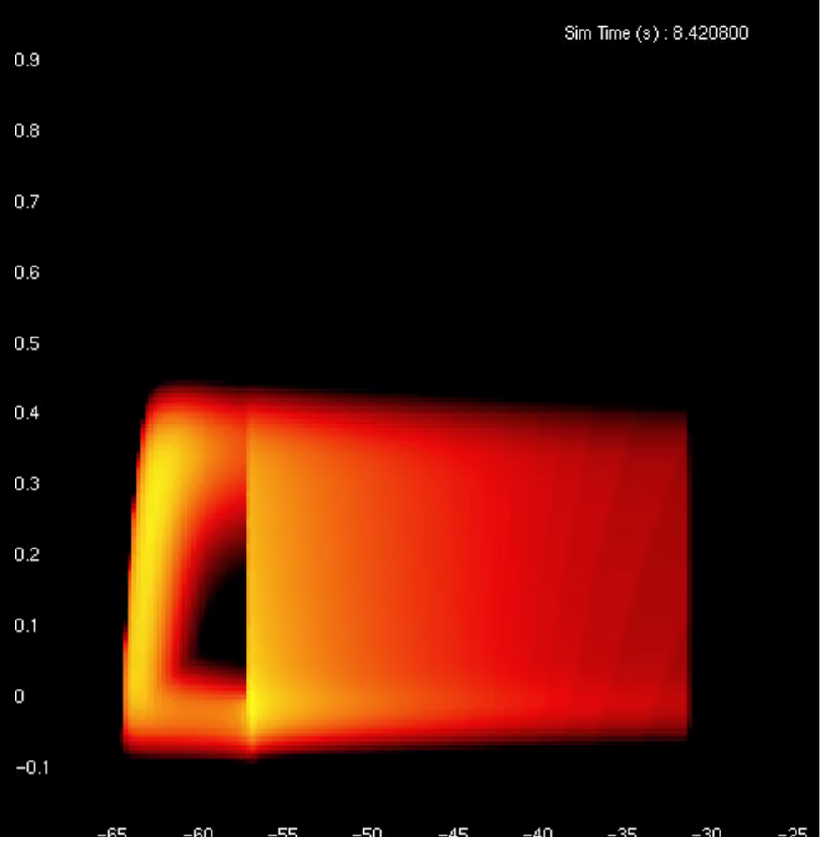}
  \includegraphics[width=0.4\columnwidth]{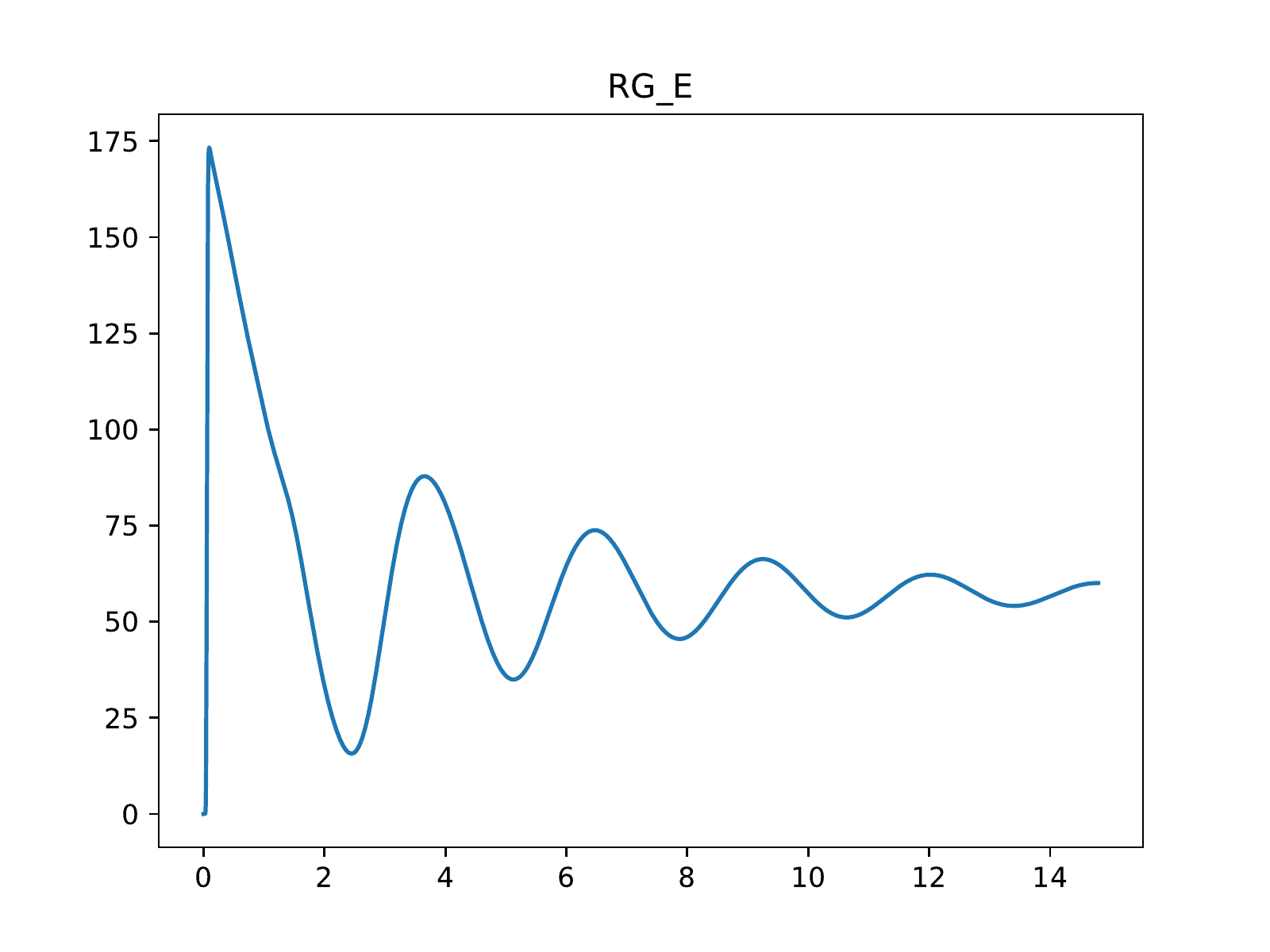}
  \caption{(left) The density plot for a population of reduced persistent sodium bursting neurons. (right) The average firing rate for the same population.}
  \label{fig:fig3}
\end{figure}

Files:\\
$PersistentSodium2D/pnap.xml$\\
$PersistentSodium2D/pnap.model$\\
$PersistentSodium2D/pnap.tmat$

\section{Reduced 3D Hodgkin-Huxley}
A 3 dimensional Hodgkin-Huxley style neuron with a time independent sodium activation. Used in \cite{rybak2006modelling} for spinal interneurons. No threshold reset mechanism is used.
 
\begin{align}
    V' = -g_{k}m_{k}^4 ( V - E_{k} ) - g_{Na}m_{\infty Na}^3h_{Na} ( V - E_{Na} ) - g_{l} ( V - E_{l} ) + I \nonumber
\end{align}
\begin{align}
    \tau_{Na}h_{Na}' = h_{\infty Na} - h_{Na} \nonumber
\end{align}
\begin{align}
    \tau_{k}m_{k}' = m_{\infty k} - m_{k} \nonumber
\end{align}

$g_{k}$, $g_{Na}$, and $g_{l}$ are constant conductances with $m_{\infty k}$, $m_{\infty Na}$, and $h_{\infty Na}$ representing typical sinusoidal curves.

\begin{figure}[H]
  \centering
  \includegraphics[width=0.4\columnwidth]{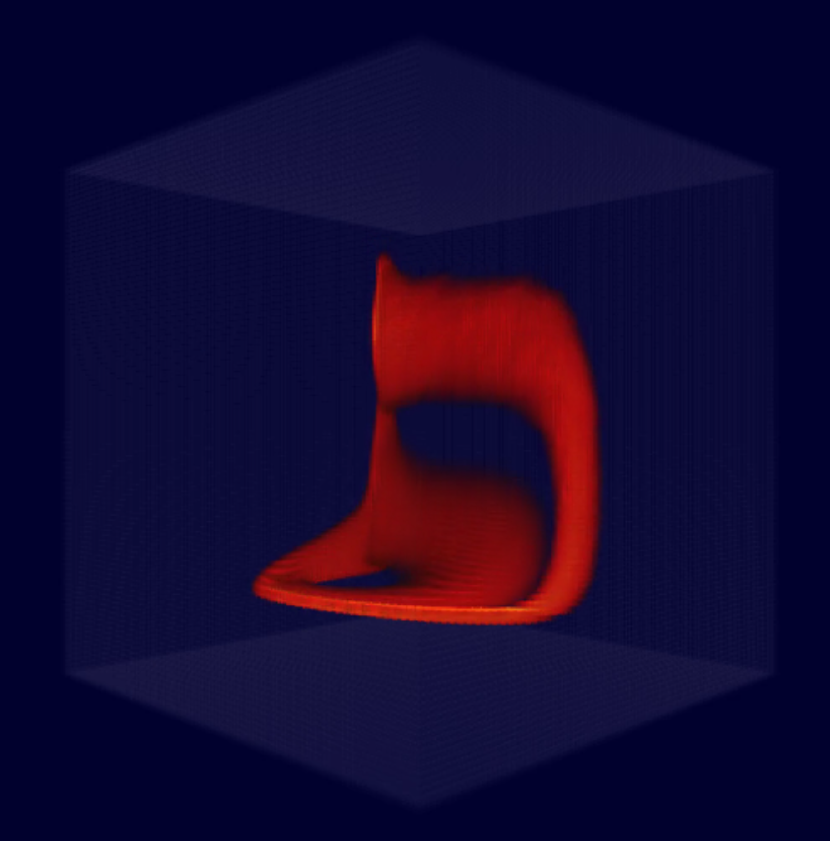}
  \caption{The 3D density plot for a population of reduced Hodgkin-Huxley neurons.}
  \label{fig:fig4}
\end{figure}

Files:\\
$Unsupported\_3D/HodgkinHuxley3D/hh\_redux.xml$\\
$Unsupported\_3D/HodgkinHuxley3D/hh\_redux.model$\\
$Unsupported\_3D/HodgkinHuxley3D/hh\_redux.tmat$

\section{2D Conductance Based Synapse}
A simple integrate and fire model coupled with a conductance based synapse. The second time dependent variable tracks the openness of the synapse. Membrane potential is driven up based on the second variable instead of direct jumps in membrane potential.

\begin{align}
    \tau V' = - g_{l} ( V - E_{v} ) - W ( V - E_{e} ) \nonumber
\end{align}
\begin{align}
    \tau_{w} W' = - W + I\nonumber
\end{align}

\begin{figure}[H]
  \centering
  \includegraphics[width=0.3\columnwidth]{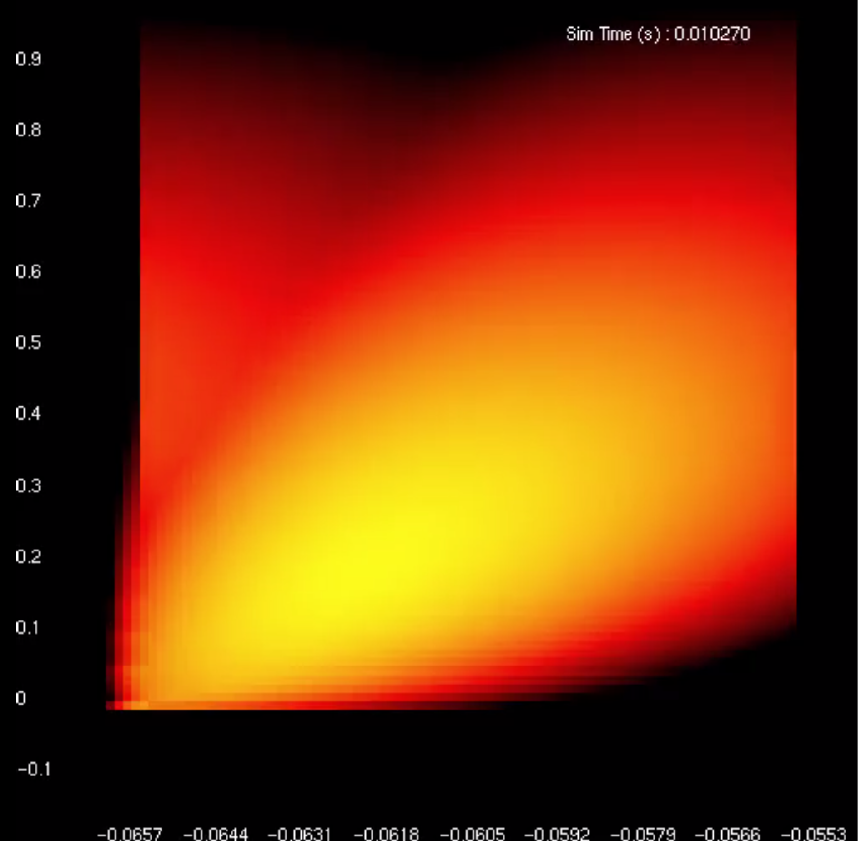}
  \caption{Density plot for a population of conductance based neurons simulated in MIIND. }
  \label{fig:fig7}
\end{figure}

Files:\\
$Conductance2D/cond.xml$\\
$Conductance2D/cond.model$\\
$Conductance2D/cond\_0\_0\_0\_0\_.tmat$

\section{3D Conductance Based Synapse}
As with the 2D version except with allowance for a second synapse, typically with a very different timescale (for example one excitatory and one inhibitory synapse).

\begin{align}
    \tau V' = - g_{l} ( V - E_{v} ) - W ( V - E_{e} ) - U ( V - E_{i} ) \nonumber
\end{align}
\begin{align}
    \tau_{w} W' = - W + I_{w} \nonumber
\end{align}
\begin{align}
    \tau_{u} U' = - U + I_{u} \nonumber
\end{align}

\begin{figure}[H]
  \centering
  \includegraphics[width=0.3\columnwidth]{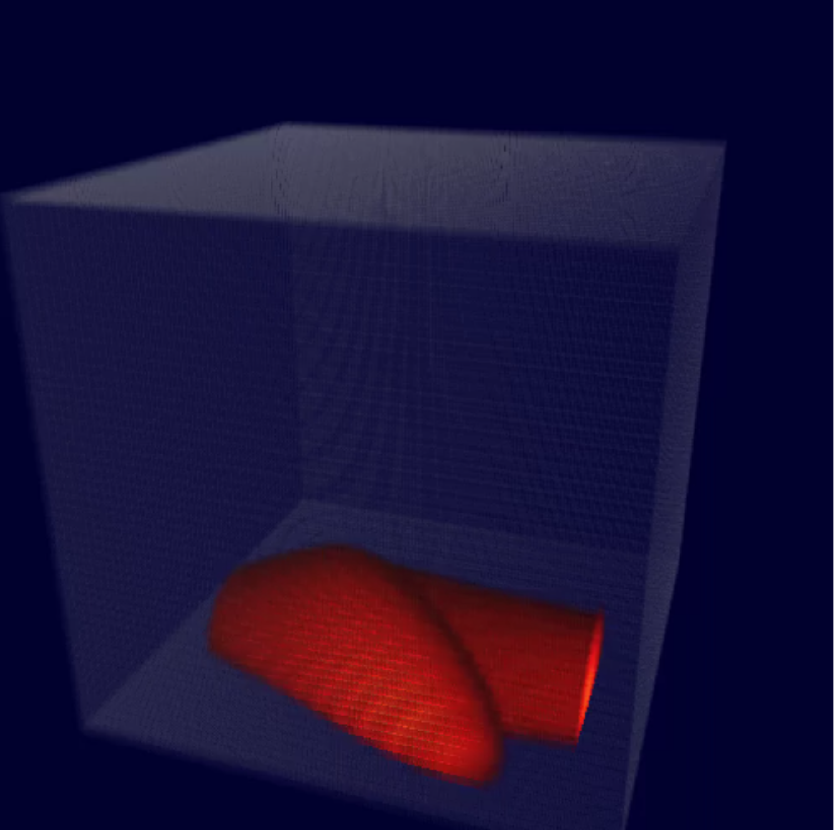}
  \caption{Density plot for a population of 3D conductance based neurons simulated in MIIND. }
  \label{fig:fig8}
\end{figure}

Files:\\
$Unsupported\_3D/Conductance3D/cond\_3D.xml$\\
$Unsupported\_3D/Conductance3D/cond\_3D.model$

\section{Hindmarsh-Rose}
A commonly used 3D model \cite{hindmarsh1984model} for exploring a wide range of bifurcations and neuron behaviours. Two variables capture the dynamics required for a hopf bifurcation from stationary to spiking on a limit cycle. The third variable represents a slow mechanism (such as persistent sodium) for controlling excitability to capture behaviours such as bursting.

\begin{align}
    X' = Y - aX^3 + bX^2 - Z + I \nonumber
\end{align}
\begin{align}
    Y' = c - d X^2 - Y \nonumber
\end{align}
\begin{align}
    Z' = r ( s (X - X_{r}) - Z ) \nonumber
\end{align}

\begin{figure}[H]
  \centering
  \includegraphics[width=0.3\columnwidth]{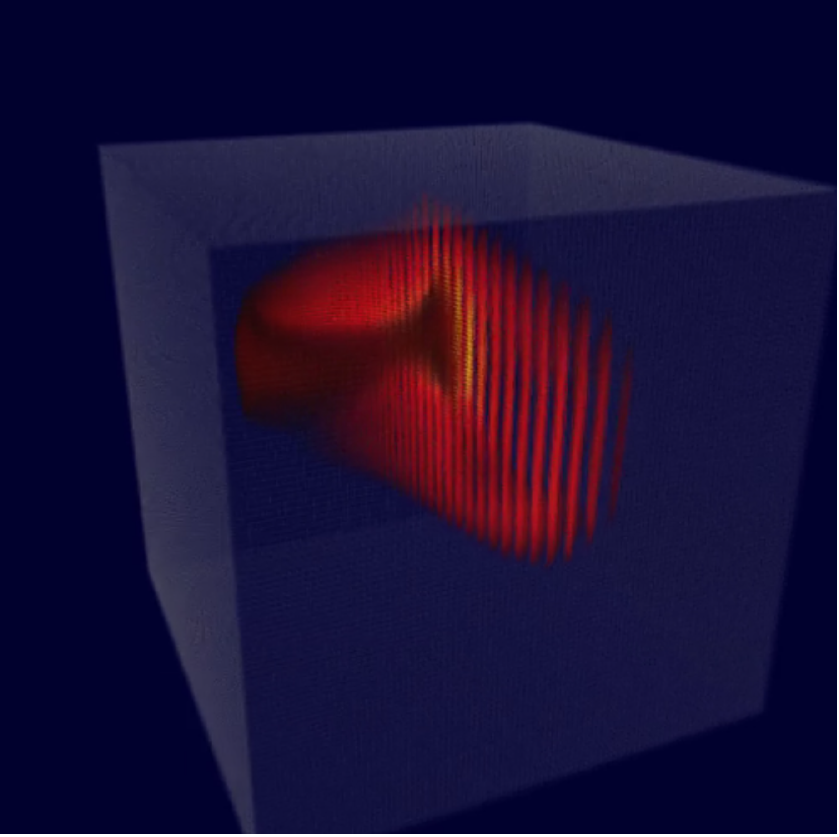}
  \caption{Density plot for a population of Hindmarsh-Rose neurons in a bursting regime simulated with MIIND. }
  \label{fig:fig9}
\end{figure}

Files:\\
$Unsupported\_3D/HindmarshRose/hr.xml$\\
$Unsupported\_3D/HindmarshRose/hr.model$

\section{Minimal Two Compartmental Motor Neuron}
A minimal two compartment motor neuron model \cite{booth1995minimal} designed to capture the bistable nature of motor neurons. One compartment produces typical tonic spiking in the soma while the second compartment describes the ``on/off'' behaviour of the dendrites. In miind, the two compartments are considered as separate populations of 2 dimensional models. As a result, the soma population is dependent on the average value of the dendrite population and vice versa. For this reason miind does not perfectly reproduce a population of motor neurons. The precise definition of the model can be found in \cite{booth1995minimal}.

\begin{figure}[H]
  \centering
  \includegraphics[width=0.3\columnwidth]{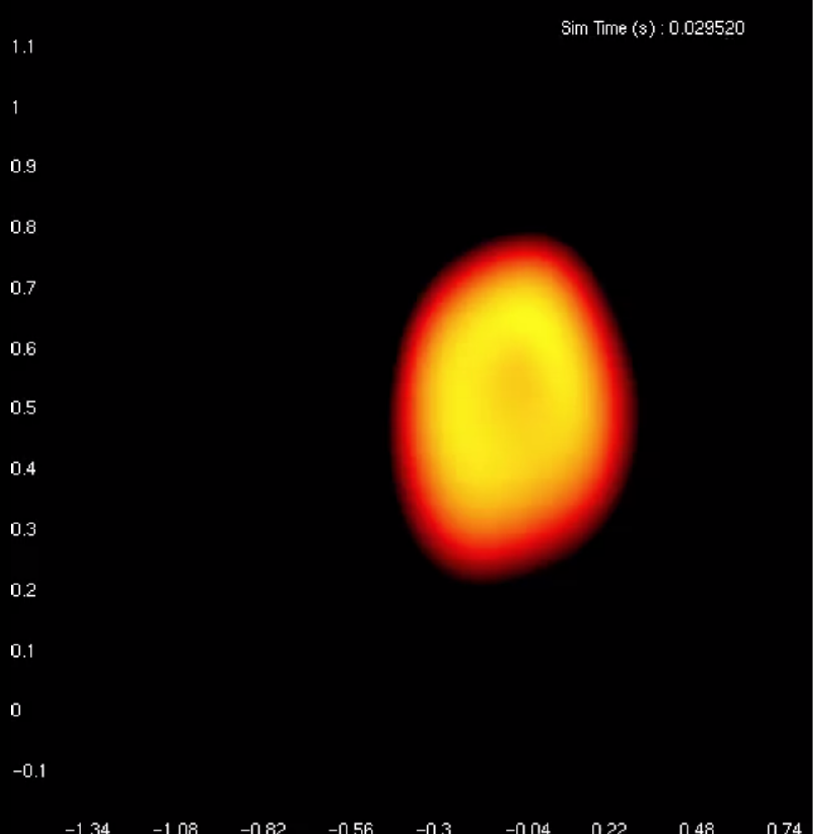}
  \includegraphics[width=0.3\columnwidth]{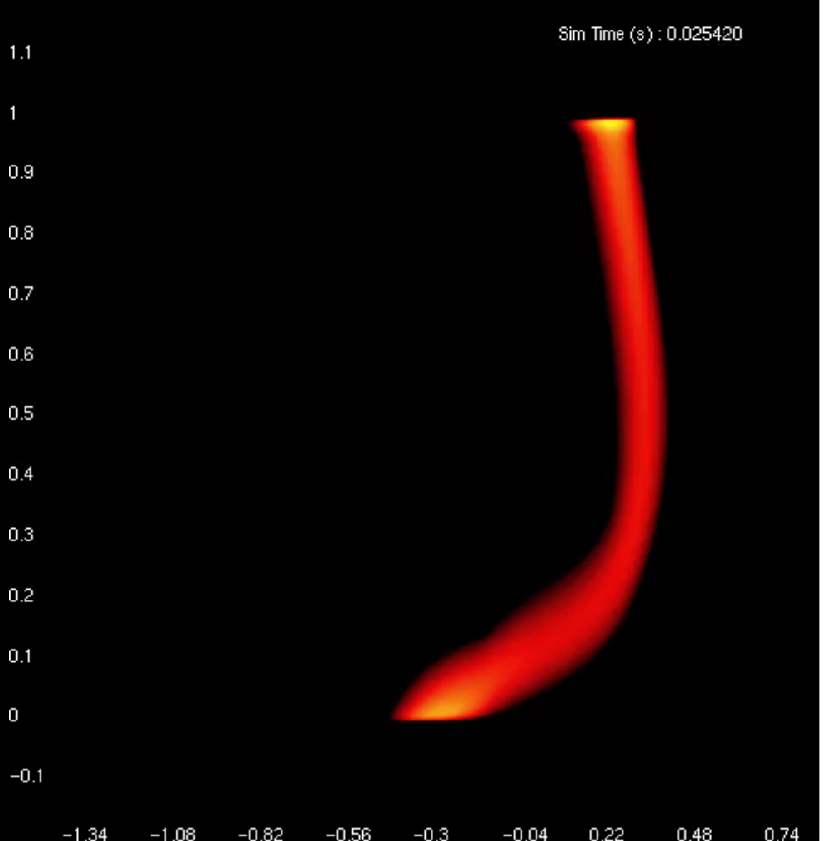}
  \includegraphics[width=0.3\columnwidth]{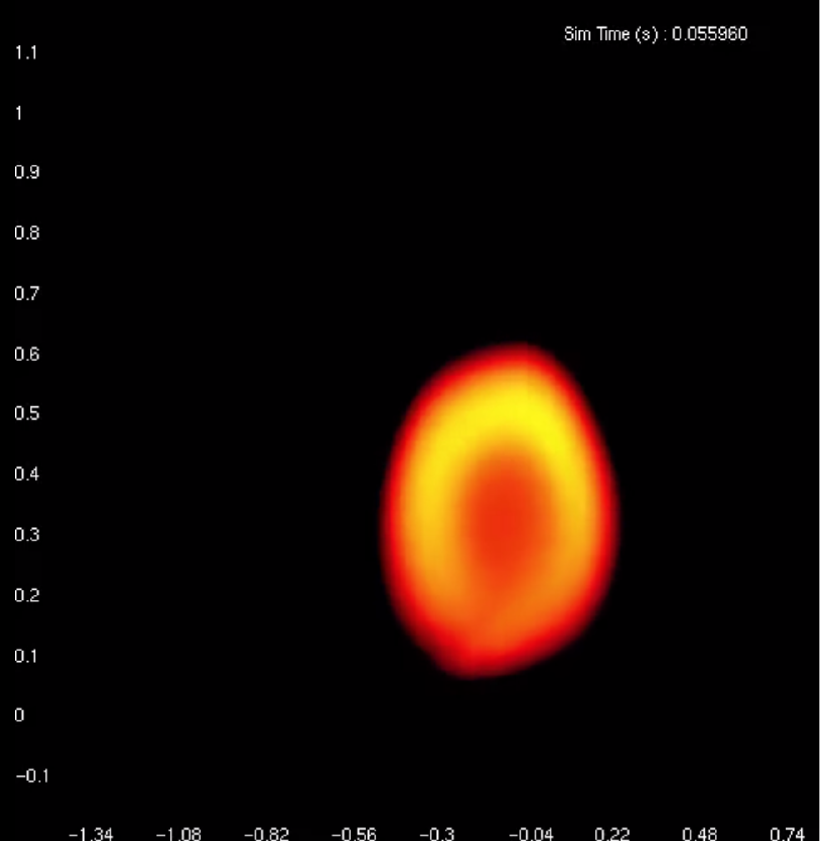}
  \includegraphics[width=0.3\columnwidth]{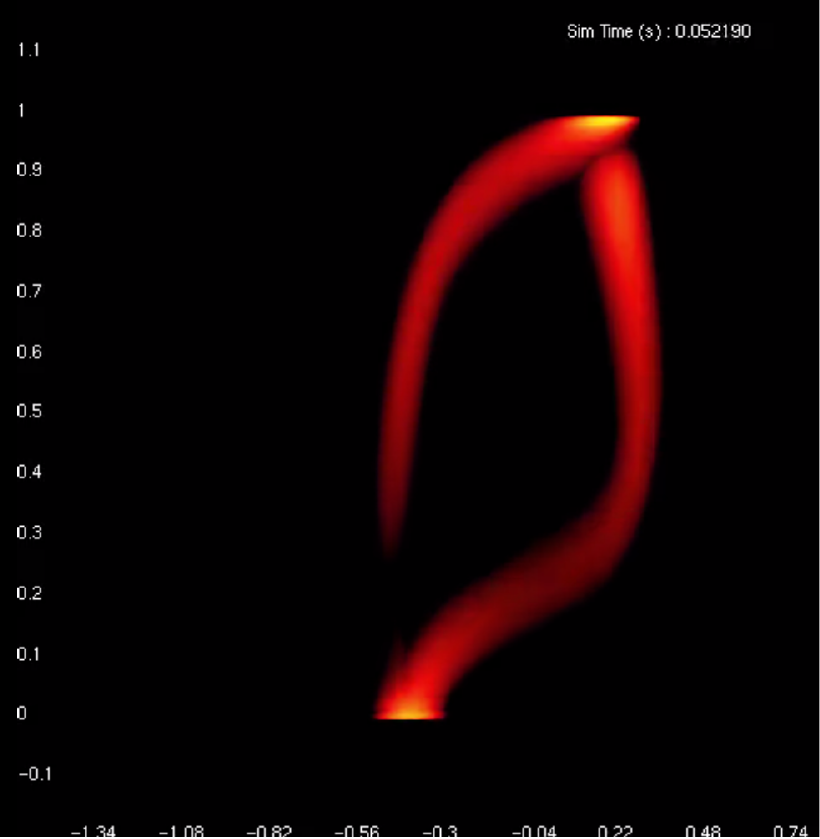}
  \caption{Density plots for two MIIND populations. The soma population (left) performs tonic spiking at two separate rates and locations in state space depending on the state of the dendrite population (right). The two populations communicate their average membrane potential to each other and a linear interpolation is made based on the desired level of soma/dendrite coupling.}
  \label{fig:fig10}
\end{figure}

Files:\\
$MinimalMotorneuron/mn.xml$\\
$MinimalMotorneuron/mn\_soma.model$\\
$MinimalMotorneuron/mn\_soma\_0\_0\_0\_0\_.tmat$\\
$MinimalMotorneuron/mn\_dendrite.model$\\
$MinimalMotorneuron/mn\_dendrite\_0\_0\_0\_0\_.tmat$

\section{2D Epileptor}
The 2D reduction \cite{proix2014permittivity} of the Epileptor model used in The Virtual Brain \cite{jirsa2014nature}. The model is parameterised to define an epileptogenic neuron vs a non-epileptogenic one. When tightly coupled together via membrane potential (instead of firing rate), an epileptogenic neuron can induce the ictal behaviour in a non-epileptogenic neuron. In MIIND, when two populations of the differently parameterised neurons are connected together via the average membrane potential of each population, the ictal behaviour can be induced in the same manner.

\begin{figure}[H]
  \centering
  \includegraphics[width=0.3\columnwidth]{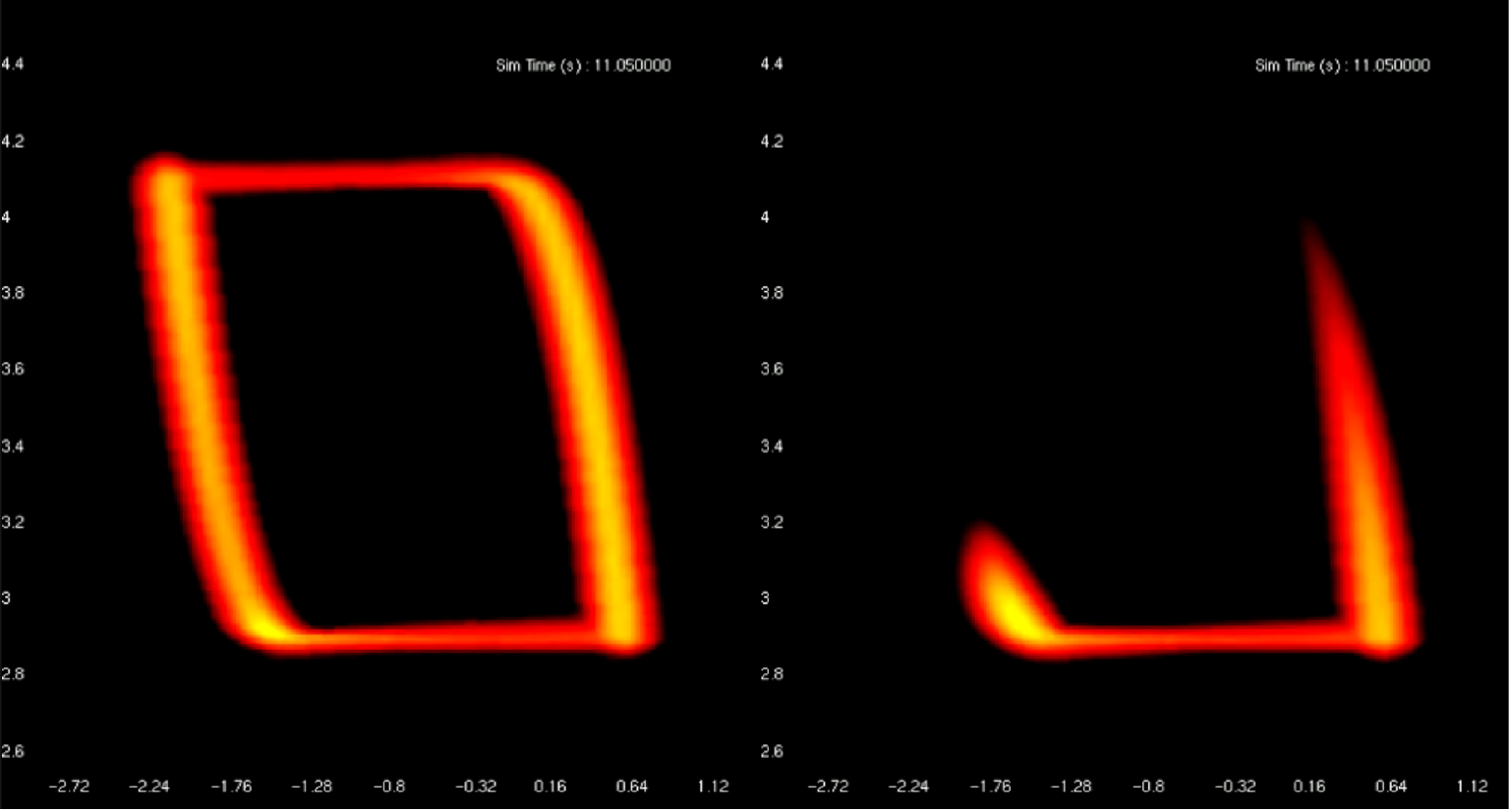}
  \includegraphics[width=0.3\columnwidth]{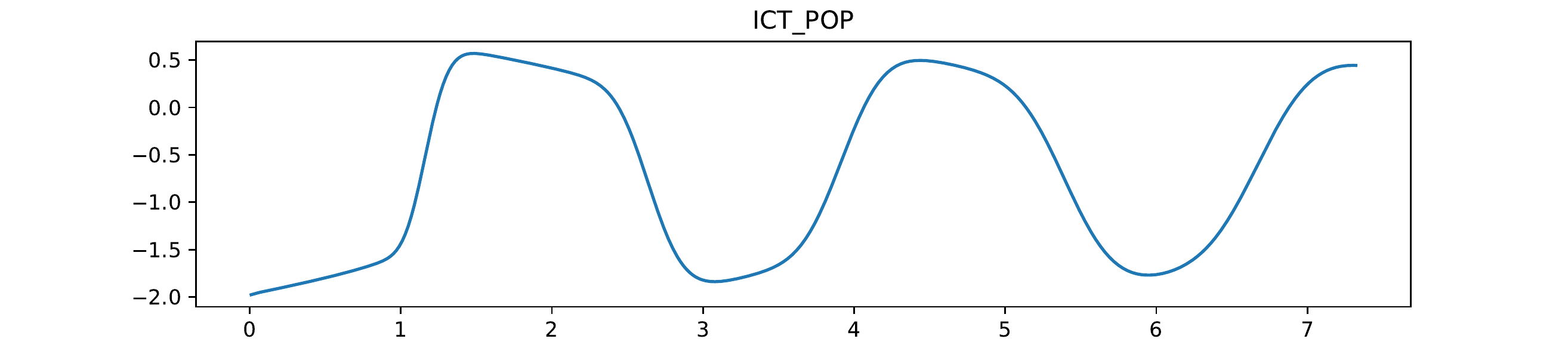}
  \includegraphics[width=0.3\columnwidth]{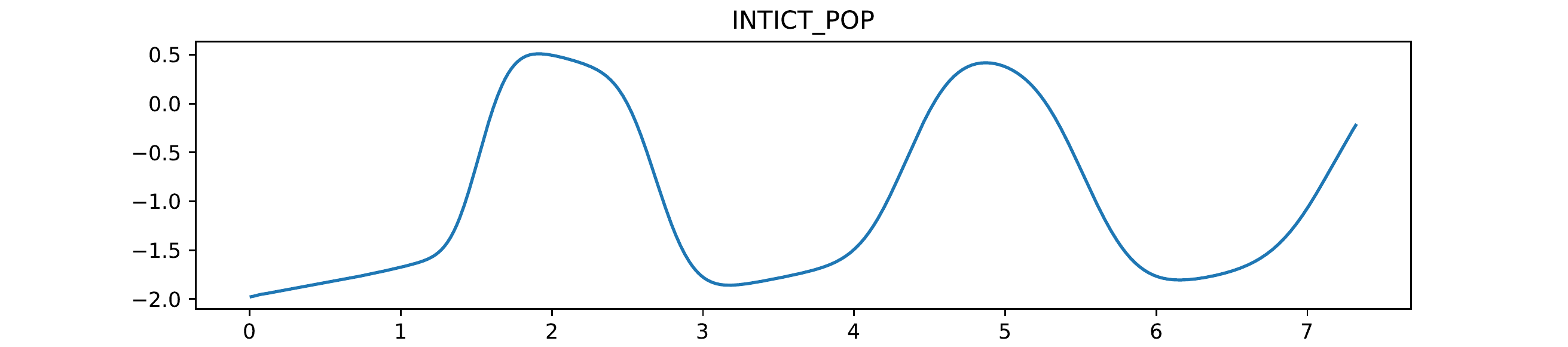}
  \caption{(above) Density plot for two Epileptor2D populations in MIIND (epileptogenic left, non-epileptogenic right). As with the motor neuron model, the populations pass a linearly interpolated average membrane potential to each other based on the coupling strength. When tightly coupled, the population on the right which would normally stay close to a quiescent stable point, is pushed into the ictal state by the other population. The middle plot shows the average membrane potential of the epileptogenic population. The lower plot shows the average membrane potential of the non-epileptogenic population which lags behind.}
  \label{fig:fig11}
\end{figure}

Files:\\
$Epileptor/epi.xml$\\
$Epileptor/epileptor\_ict.model$\\
$Epileptor/epileptor\_ict\_0\_0\_0\_0\_.tmat$\\
$Epileptor/epileptor\_intict.model$\\
$Epileptor/epileptor\_intict\_0\_0\_0\_0\_.tmat$

\section*{Acknowledgements}

 This project received funding from the European Union’s Horizon 2020 research and innovation programme under Specific Grant Agreement No. 785907 (Human Brain Project SGA2). The funders had no role in study design, data collection and analysis, decision to publish, or preparation of the manuscript.

\medskip
\printbibliography

\end{document}